# A Survey on Grading Format of Automated Grading Tools for Programming Assignments

**Aditi Agrawal, Benjamin Reed**

*San Jose State University(USA)*

## Abstract

The prevalence of online platforms and studies has generated the demand for automated grading tools, and as a result, there are plenty in the market. Such tools are developed to grade coding assignments quickly, accurately, and effortlessly. Since there are varieties of tools available to cater to the diverse options of programming languages and concepts, it is overwhelming for any instructor to decide which one suits his/her's requirements.

There are several surveys studying the tools and giving insights into how they function and what they support. However other than knowing the functionality, it is important for an instructor to know how the assignments are graded and what is the format of the test cases. This is crucial since the instructor has to design the grading format and therefore requires a learning curve. This survey studies and evaluates the automated grading tools based on their evaluation format. This in turn helps a reader in deciding which tool to choose and provides an insight into what are the assessment settings and approaches used in grading the coding assignment in any specific grading tool.

## 1. Introduction

There is a shift to virtual platforms in academia due to advancements in computer hardware, better internet connectivity, and the onslaught of the pandemic. Grading programming assignments on an online platform is laborious, time-consuming, and error-prone. As a result, numerous automated tools are being used by many institutions to grade coding assignments.

Grading a coding assignment requires thorough evaluation. The code must work for all the possible scenarios, with optimized time and space complexity. It must be error-free, conforming to the problem specification and solving it with the correct approach. Automated grading tools can perform many of these tasks to grade the coding assignment and provide quick feedback.

These tools check for programming errors, plagiarism, coding style, code design, output comparison, and structural similarity.There are numerous surveys studying and evaluating the automated grading tools on these features. They compare which of the above features are supported by which tool, but they do not do a comparison of how the evaluation criteria of the assignments are specified by the instructor.

This paper studies and compares the different automated tools and their specification formats. Some tools use a GUI to author the grading format of the assignment while others use specification files with test cases and other grading parameters.The authoring interfaces and specification formats of these tools enable a set of criteria to be used in evaluating the submission.This knowledge of formats, enables an instructor/grader to choose a tool and edit the grading format as per their assignment requirement while using that specific tool. This survey provides readers with a systematic comparison of the expressibility and compatibility of the available automated grading tools. Section 2 covers the work already done in this field followed by Section 3 which analyses each tool. The results and conclusion after studying the tools are covered in Sections 4 and 5 respectively.

## 2.  Related Work

Among the surveys on automated grading tools for coding assignments available today, most segregated and compared the tools in the areas such as programming languages, and dynamic and static analysis.

K.A.Mukta in [1], surveyed the automated grading tools that were in the market until 2005. She studied and outlined the features that could be automatically evaluated by these programming tools. Some of the features pointed out were, the usage of a sandbox environment for security purposes, functionality, and efficiency of the code by testing it against defined test cases, catching programming errors such as runtime and compilation errors, running student-defined test cases, coding style, programming language specific issues such as memory leak check for C, and special features (certain libraries are used or not).

Building on similar work, P. Ihantola et al. [2] and Lian et al. [3] covered the automated grading tools until 2010. Other than evaluating them on the same features as done in [1], [2] has also compared the tools based on the platforms they are offered on, such as whether the tool is a standalone tool or integrated with LMS such as CodEval[4], AutoGrader [5], and on resubmission criteria which includes limiting the number of submissions, penalty on late submission, limiting the amount of feedback provided for instance in [4], and hybrid approaches such as in Marmoset[6], where the limit of submissions is optional and set by the instructor.

Similar to the surveys mentioned above, the authors in [7], evaluate the tools until 2021 on additional criteria such as CS domains supported, and the information offered to an instructor to optimize the learning experience for students. They extensively segregated the tools into CS domains such as visual programming, System Administration(LINSIM[8]), formal language, and automata(JFLAP[9]), web development, and software modeling by comparing class diagrams, and other aspects of UML, parallel computing(SAUCE[10]), and software testing which includes evaluating the code base on test cases.

Apart from evaluating the technical differences and concepts, some surveys evaluated the tools based on the pedagogical aspect. [1],[11],[12], have compared and contrasted grading tools on automatic assessment(GAME[13]), semi-automatic assessment(JACKSON[14]), and hybrid assessment where some static and dynamic analysis is done by the tool and the qualitative analysis such as code style, and which grade to give out is done by the instructor or the grader.(Webcat[15], MOSSHAK[16]).

Another criterion used in [1] was distinguishing the tools based on the formative versus summative approach. The author describes the formative approach as the one where the student can submit the assignment multiple times thereby helping the student improve the programming skills as done in CourseMaker[17], and Codeval [4]. In contrast, in a summative approach, the student is allowed to submit the code only once, for instance in BOSS[18,19].

Other than the technical and pedagogical evaluation, there are surveys on varied aspects of automated grading tools. In [20], J.R. Alamo, conducted an experiment on his students by introducing various online assessment tools(Programmr[21], Code Step by Step [22]) every fall semester and gathered the feedback from students to conclude how these tools had helped them in improving the grades.

An intriguing aspect of automated tools was studied in [23] by R. Queiros and J.P. leal, where they evaluated the tools based on their interoperability ability. They took 15 tools and studied how easily these tools interoperate in terms of programming exercises, users, and assessment results. They concluded that more than 50 percent of these tools lacked sophisticated interoperability and therefore there is a need to focus on this segment.Though there are good literature and surveys available on automated grading tools, none have studied in-depth the assessment specification of the tools in the market.

## 3. Assessment Of Tools

### 3.1. CodeEval

CodEval[4] is an automated evaluating tool for coding assignments developed at San Jose State University, integrated with Canvas LMS, and supports Java and C programming assignments. The software has its own specification format which has to be used by the instructor to design the test cases and other checks required for the assignment. When a student submits the assignment on Canvas, the Canvas API interacts with the SJSU web server and sends the submission to it, where the CodEval instance evaluates the output of the student's submission against the test cases mentioned in the specification file. An example is mentioned in figure 1, for the complete format please refer to [4].

```
RUN evaluate.sh
C javac edu/sjsu/CS001/Hello.java

T java edu.sjsu.CS001.Hello ben
X 0
O Hello ben
```

**Fig 1: Snippet from specification file of CodEval.**

### 3.2. Cloud Coder

Cloud Coder[24], is an open source web-based automated assignment grading utility for C/C++, Java, Python, and Ruby. It is an on-premise software installed on Linux Operating System, where the instructor can create a course, register the students, create exercises and evaluate the programming assignments. The instructor can choose the type of programming language to be used by the student and the type of problem while creating the exercise. The various types supported are C_PROGRAM, JAVA_PROGRAM(complete program) and C_METHOD, JAVA_METHOD, PYTHON_METHOD, and RUBY_METHOD where only a method has to be implemented and tested.

There is an "Add test case button" where the user can enter the test case for the problem. These three input fields here namely, Test case name, Test input, and Test output. In Test Input, the instructor can enter the inputs to be passed to the code. While passing the test inputs to methods, it will be a list of literal arguments whereas for the program it will be a string sent to the standard input. In Test Output, the user has to add regular expressions for program-type problems so as to compare the output from the submission to the regular expression. In case of method type problems, the user has to enter the literal values of expected output, which will be compared with the output of the submitted code using equality operators and functions such as "==" in C++, C. Python, and Ruby and the '*equal*' method in Java.

### 3.3. ASSYST

Assyst[25] stands for Assessment System, an on-premise software developed at the University of Liverpool, UK, and was used initially for grading Ada language assignments. The student using the graphical interface can submit the problem which is first checked for compilation and linking errors. Once passed, the output of the submitted code is compared with the output specification defined by the tutor and with the test data passed by the student in the software. The technique used to analyze the output is pattern matching. ASSYST uses Unix Lex and Yacc tools to generate the program which compares whether the student's output matches the output in the output specification file.

### 3.4. MarkUs

MarkUs is a web-based application for Ubuntu systems, developed using Ruby. The languages supported are Java, Haskell, Python, and Ruby. In order to test the output of the student's code, it uses test code that is designed by the instructor. The test code is uploaded in the particular folder mentioned in [26] and is called when the student submits the code. The instructor needs to select the Tester type which is the testing framework that is required to evaluate the problem. There are various testers available.

- Python which uses *pytest* or *unittest* frameworks to perform the testing for python submission. An example of how the test cases are mentioned in *pytest* and *unittest* is as follows:

```
import unittest
class TestStringMethods(unittest.TestCase):

    def test_upper(self):
        self.assertEqual('foo'.upper(), 'FOO')

    def test_isupper(self):
        self.assertTrue('FOO'.isupper())
        self.assertFalse('Foo'.isupper())

    def test_split(self):
        s = 'hello world'
        self.assertEqual(s.split(), ['hello', 'world'])
        # check that s.split fails when the separator is not a string
        with self.assertRaises(TypeError):
            s.split(2)
if __name__ == '__main__':
    unittest.main()
```

```
# content of test_assert1.py
def f():
    return 3
def test_function():
    assert f() == 4
```

**Fig 2: Example of Unittest file**    **Fig 3: Example of Pytest test file**

- Java uses Junit test code to define the test cases.
- Pyta can also be used to test python code. This acts as a wrapper around *pylint* and is used to perform static analysis of the code.
- Haskell tester uses Quickcheck [27], which is a library to test program properties.

### 3.5. ECSpooler

ECSpooler [28] is integrated with the content management system(CMS) Plone [29] to provide tutorials, slides, and automated assessment of programming assignments for Java and Haskell. The structure is three-tier, where the first tier is the front end, which in this case is the user interface of Plone. The second tier contains the spooler which collects students' submissions, corresponding unit tests, and chosen backend and sends it to the appropriate Backend. The chosen Backend calls the compiler and/or interpreter on the submitted code. Once passed, it moves on to run the submitted code against the test code passed by the spooler. The feedback on failed and passed test cases is sent back to the user. The unit test cases are entered in the ECAutoAssessmentBox[28] in the format of JUnit test cases for Java problems and Quickcheck specifications for Haskell problems. Other than a unit test, model solutions provided by the instructor are also compared.

### 3.6. CodeCover

CodeCover [30], is integrated with Moodle, using Java, and supports Java assignments. CodeCover takes the student's submission (software under test), and JUnit test cases as input, evaluates the submission against the test cases and sends back the feedback in the CSV and HTML format to Moodle to display it to the user. The Moodle interface provides text fields to enter the source code and student's test data which are sent to Codecover. The format in which the test data has to be entered is specified on the screen for the student.

## 3.7.  CodeAssesor

CodeAssessor [31], is a web-based standalone tool that supports C++ programming assignments. It uses a MySQL database to store students' submissions, problem statements, and student's grades. The interface provides text boxes for the instructor to enter the problem name, description, model solution, and test cases. The format of the specification is plain text. After entering all the details, the instructor can select the "Publish Problem" button to publish the assignment. The student is also provided with the text box to enter his/her code and test parameters. Once the student submits the code, the student's code and the instructor's code are compiled using the GNU compiler and tested against the test case provided by the instructor and the student. The number of test cases passed or failed is stored in the database and then displayed to the user.

## 3.8.  JavaAssess

JavaAsses [32], is a Java library with API that includes 200 methods to evaluate a Java assignment's correctness. It can do both black box(output comparison) and white box(internal properties of the code) comparisons. JavaAssess is not an assessment system but a library that can be used to evaluate Java code. The tool has four modules namely Introspector, Intercessor, and Tester, and the fourth module internally calls these three. In order to use the fourth module, the programmer's class must extend the Assessor class in *codeassess.javaassessment* package. The instructor will have to design a java code that makes use of the JavaAssess library methods and compares the output generated by the methods to the desired output mentioned in the code by the instructor to grade the assignment.

## 3.9.  JACK

Jack [33], is a web-based automated grading system for Java assignments that performs both static and dynamic analysis of submitted code. As part of static analysis, the *java2ggx* tool is used to convert the source code into a graph structure in *ggx* file format used in Abstract Graph Grammer(AGG) [34]. After which Control tool is used to choose the rules that need to be applied to the graph to check its correctness. The rule contains three-part, LHS, Negative application condition(NAC), and RHS, where firstly the checker would try to match the LHS to the graph generated from the source code. Once a match is found one or more NAC is applied and if the condition for the NAC is satisfied then it performs nothing and moves to the next rule, whereas if the NAC condition is satisfied or not present at all, the nodes and edges mentioned in LHS are replaced with the ones present in RHS.

As part of dynamic analysis, the submitted code is tested against the input parameters, which are designed in the form of unit test cases, where a method is called on the source code and the return value is compared with the reference value. The unit tests are written in Java following the approach mentioned in [35].

## 3.10.  CodeLab

CodeLab [36], is a web-based programming assignment assessment tool, which supports Python, Java, C, C++, and C#. It has a repository of programming problems for an instructor to choose from while creating a course and assignments in it. On the other hand, if the instructor wants to customize a problem, then the instructor can click on the "Edit Exercise" option, where the interface has multiple text boxes, such as specification type(method, program), Instructions(HTML), Solutions(instructor's code), Test Case Specification and so on and so forth.

The Test Case Specification follows a two-column format, where the left-hand side of the statement contains the input which needs to be sent to the program, followed by the "->" symbol, followed by the expected output from the executed program. Each line in the specification text box represents one test case. Other than standard input and output, files can also be passed as input and output, where the filenames should match the ones mentioned in the source code. An example is presented in Figure 4.

```
1  ($stdin, $file("Operand1") -> $stdout, $file("Add.out"), $file("Subtract.out"), $file("Multiply.out"), $file("Divide.out") )
2  ("20\n", "40" -> "fileOperand=40\n","60","20","800","2")
3
```

**Fig 4: Test case specification format.**

Here the first line is the first test case, where the files mentioned on the right side of the arrow are the output files and "Operand1" is the input file fed to the source code. In the next line, the other test case takes literal values as input. Another scenario of a test case is when links are created. Each link contains the multiple-line input fed to the code and the corresponding output link with multiple-line output. An example of such a specification is shown in Figure 5.

```
1  ($stdin -> $stdout)
2  ("abc def abcdefghijk p\nmoo oink\n\n\nshuffleboard preponderance\n" -> "abcdefghijk\nshuffleboard\npreponderance\n")
3  ($link("input1") -> $link("output1"))
4  ($link("input2") -> $link("output2"))
5  ($link("input2") -> $link("output2"))
6
```

Links: output1, output2, input2, input1, +

Name: output1

```
1  abcdefghijk
2  shuffleboard
3  preponderance
```

**Fig 5: Test case specification with links.**

## 3.11. Kattis

Kattis is an online platform for programming assignments and contests that supports C, C++, Java, Python, and Rust. The test cases are stored in /data/sample and /data/secret folders where for each test case there will be two files, one for the input and one for the expected output. There is also the concept of input validators, which validates the format of the input files. There is *problem.yaml* file which contains metadata such as license, and *problem.tex* file which contains the description of the problem. Figure 6 depicts the input and expected output files and Figure 7 depicts the *problem.tex* file.

```
10 12
71293781758123 72784
1 12345677654321
```

```
2
71293781685339
12345677654320
```

**Fig 6: An Input file and expected output file**

```
\problemname{A Different Problem}

Write a program that computes the difference between non-negative integers.

\section*{Input}

The input consists of multiple test cases (at least $1$ and at most
$40$), one per line.  Each test case consists of a pair of integers,
between $0$ and $10^{15}$ (inclusive).  The input is terminated by
end of file.

\section*{Output}

For each pair of integers in the input, output one line, containing the absolute
```

**Fig 7: Example of problem.tex file.**

### 3.12. Submitty

Submitty[37] is an open-source automated grading tool for programming assignments and exams, supporting Python, C/C++, Java, Scheme, and Prolog assignments. It performs black-box testing, output comparison, static analysis, and memory leak check(Dr Memory or Valgrind). The specification format followed by Submitty is in the form of key-value pairs separated by colon inside curly braces. The left-hand side of the colon depicts the different features required by the code such as test cases, grades associated with the assignment, and title(which language is used). The right-hand side consists of the value for the corresponding key. Submitty tool will follow the specification file to read the inputs, outputs, and reference output for the program. It uses the Linux tool *"diff"*, to compare the student's output and expected output, Junit test cases to validate Java code, and equality operators such as "==", *"ge"*, and *" le"* in the case of literal comparison. An example of the specification file is shown in Figure 8.

```
{
  // For compiled languages, typically two testcases are used to allow po
  // to be asssigned independently for compilation and execution.
  "testcases" : [
    {
      // Indicate that this is a compilation step.
      "type" : "Compilation",
      "title" : "C++ - Compilation",
      "command" : "clang++ -Wall -o a.out -- *.cpp",
      // Name of the result of compilation.
      "executable_name" : "a.out",
      // Point value of compilation.
      "points" : 5
    },
    {
      "title" : "C++ - Execution",
      "command" : "./a.out",
      // Point value of correct output.
      "points" : 15,
      "validation" : [
        {
          "method" : "myersDiffbyLinebyChar",
          "actual_file" : "STDOUT.txt",
          "description" : "Program Output",
          "expected_file" : "test1_output.txt"
        }
      ]
    }
  ]
}
```

**Fig 8: Specification format for Submitty.**

### 4. Result and Discussion

It is evident from the previous section that there are numerous ways an instructor can specify assessment specifications. For some tools, the instructor can use the pre-existing knowledge of a language such as Java to design the test cases for example in Codecover[30] whereas other tools require the instructor to learn the specific format, for example, Codeval[4]. In order to sum up the above-mentioned tools, the following table is created to show the tools and their specification formats for grading assignments.

| Tool | Specification Format/Evaluation technique |
|---|---|
| CodeEval | CodeEval specific Format, Unix tool *diff* |
| Cloud Coder | Regular expression, Equality Operators |
| ASSYST | Plain text test cases, Lex and Yacc |
| MarkUs | Pytest, Junit, Unittest, Quickcheck |
| ECSpooler | Junit format |
| CodeCover | Junit format |
| CodeAssesor | Plain text test cases, Equality Operators |
| JavaAssess | Java |
| JACK | JUnit format |
| CodeLab | CodeLab specific format |
| Kattis | Plain text test cases, Latex, Equality Operators |
| Submitty | JSON format, Unix tool *diff*, Equality Operators |

**Fig 9: Table containing tools and their specification format types.**

## 5.  Conclusion

There are several automated grading tools for programming assignments. As mentioned in the introduction, there are several surveys as well, studying the various aspects of these tools. This survey studied and summarized the grading format of 12 such tools. Most of the tools specified in this paper utilize the existing test formats such as JUnit, Pytest, and Quickcheck. However, there are some tools that have introduced their own formats where some are similar to JSON, and some have an entirely new format. The analysis provided on the specification format of each of these tools should help the reader in understanding the learning curve for the tool and hence in the decision to use the tool.